\documentstyle[epsfig]{aipproc}

\def\MeV{Me\kern-0.11em V}
\def\keV{ke\kern-0.11em V}

\begin{document}

\title{The Orbital Ephemeris and X-Ray Light Curve of Cyg X-3}

\author{Steven M. Matz$^*$}
\address{$^*$Northwestern University\\
Dept. of Physics and Astronomy\\
Dearborn Observatory\\
Evanston Illinois 60208-2900\\
}

\maketitle

\begin{abstract}

The orbital dynamics of Cyg X-3 are a key to understanding this
enigmatic X-ray binary. Recent observations by the RXTE ASM and the OSSE
instrument on GRO enable us to extend the baseline of arrival time
measurements and test earlier models of orbital period evolution. We
derive new quadratic and cubic ephemerides from the soft X-ray data
(including ASM). We find a significant shift between the predicted soft
X-ray phase and the light curve phase measured by OSSE from $\sim 44$ to
130 \keV. Some of the apparent phase shift may be caused by a difference
in light curve shape.

\end{abstract}

\section*{Introduction}

Cygnus X--3 is a unique and poorly understood X-ray source which may
represent a very short-lived, transitional state of binary evolution. A
study of its dynamics (via the orbital ephemeris), and local environment
(via the light curve and spectrum) may have broad implications for X-ray
binaries in general. The period (4.8 hr, presumably orbital) is
characteristic of low mass X-ray binaries. However, there is evidence
that the companion is actually a high-mass Wolf-Rayet star
\cite{kerk92}. The mass loss from the stellar wind of such a star could
explain the large {\em positive} $\dot{P}$ measured in Cyg X--3, which
is inconsistent with mass transfer via Roche lobe overflow
\cite{moln88a}. The light curve, observed in IR and X-rays, is
asymmetric with a non-zero eclipse, resulting, in most models, from
interactions in a dense cocoon or wind around the system
\cite{prin74,milg78,whit82,will85}. If the asymmetry is instead caused
by an elliptical orbit, the light curve shape is expected to change
significantly over time due to apsidal motion \cite{ghos81} which could
also produce some or all of the apparent period increase \cite{elsn80}.
Therefore, measurements of the light curve shape are directly related to
the orbital dynamics.

\section*{Observation Summary}

OSSE \cite{john93} observed Cyg X-3 for $\sim 65$ days in 7 intervals
between 1991 May 30 and 1994  July 12. We use the 2-min
background-subtracted $\sim 50$ \keV\ to 10 \MeV\ spectra produced by the
standard OSSE spectral analysis routines for these 7 observations. There
was no significant flux contribution from Cyg X--1 in any of the source
or background pointings. Some results of the earlier OSSE observations
have been described elsewhere \cite{matz94}.

For the ASM we used the background subtracted 2--10 \keV\ Cyg X-3 counting
rates from the quick-look results provided by the ASM/RXTE team over the
WWW. The data analyzed here cover 1996 Feb 22 -- 1997 Apr 11.

\begin{table}[tbh]
\caption{ASM Arrival Times}
\label{arr_time}
\begin{tabular}{cccc}
Interval&Cycle&Arr. Time&$\sigma_{T}$\\
(HJD $-$ 2440000.)& &(HJD $-$ 2440000.)&(days)\\

10135.8--10235.4& 46251.0000  &  10185.6237 &   $1.0\times 10^{-3}$\\
10236.0--10335.9& 46753.0000  &  10285.8687 &   $1.1\times 10^{-3}$\\
10335.9--10435.8& 47254.0000  &  10385.9134 &   $1.4\times 10^{-3}$\\
10435.8--10535.7& 47754.0000  &  10485.7588 &   $1.2\times 10^{-3}$\\

\end{tabular}
\end{table}

\section*{ASM Arrival Time Derivation}

To correct for large, long term changes in the source DC flux level the
ASM 2--10 \keV\ Cyg X-3 rates were first ``cleaned'' by fitting a line to
successive 10-day segments of the data and subtracting the fit. The
linear fit removes trends on time scales much longer than an orbital
period. This process did not significantly change the arrival times
derived from the data.

The cleaned data were divided into approximately 100-day intervals. Long
intervals are needed to average over the cycle-to-cycle light curve (LC)
variations and to completely sample the Cyg X-3 orbit. The data in each
interval were then corrected to the SSB and epoch-folded (using a
constant period ephemeris) to produce orbital light curves. We fit the
resulting light curves with the EXOSAT template \cite{vand89} to
determine the relative phase of the LC minimum. This phase is then used
to calculate an arrival time for a cycle near the middle of the
observation interval \cite{matz96}. The arrival times are shown in Table
\ref{arr_time} in heliocentric JD (HJD) $-$ 2440000.

\begin{figure}[tbh]
\centerline{\epsfig{file=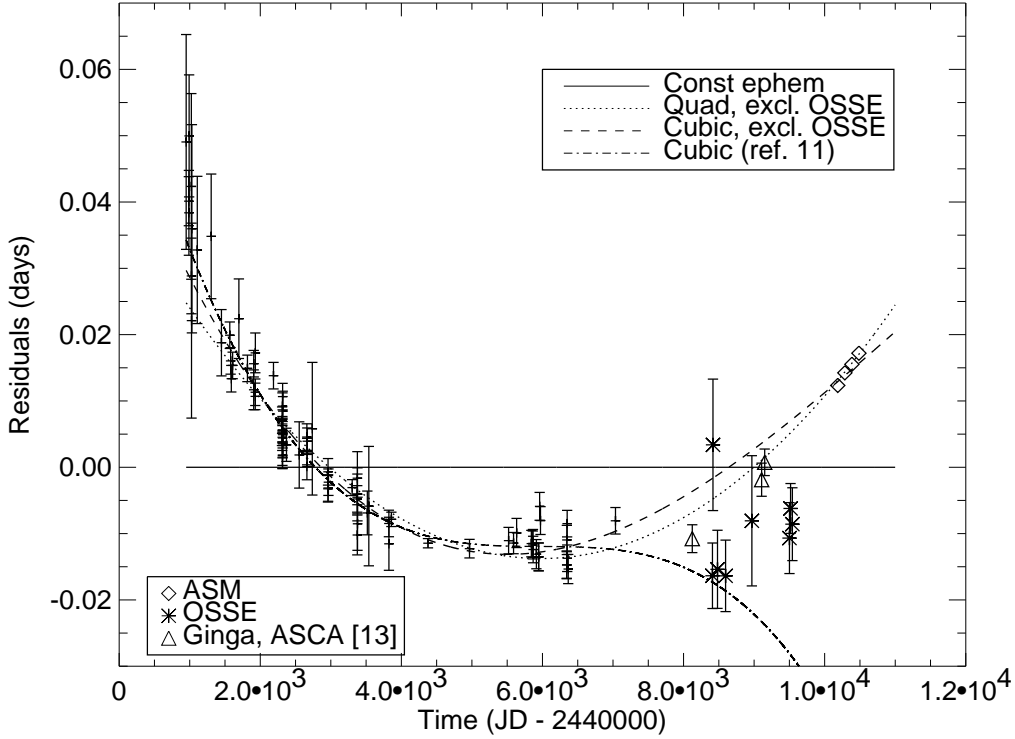,width=5.4in,angle=90}}
\caption{Arrival time residuals for OSSE, ASM, and historical data. The
errors on the ASM points are approximately equal to the symbol size.
Also shown are the best-fit quadratic and cubic ephemerides. An
extrapolation of an older cubic fit \protect{\cite{vand89}} with a
larger negative $\ddot{P}$ is shown for comparison.}

\label{res_plot} 

\end{figure}

Figure \ref{res_plot} shows differences (residuals) between the arrival
times predicted by a constant  period ephemeris and those observed by
the ASM and by other soft X-ray instruments going back to 1970
(\cite{kita95} and references therein, plus \cite{bonn85}). Overplotted
are fits with quadratic and cubic ephemeris models.

\begin{table}[tbh]
\caption{PRELIMINARY Best-Fit Quadratic and Cubic Ephemerides}
\label{ephem}
\begin{tabular}{lr|lr}
\multicolumn{2}{c|}{\( T_n = T_0 + P_0n + c_0n^2\)}&
\multicolumn{2}{c}{\(T_n = T_0 + P_0n + c_0n^2 + dn^3\)}\\
&&&\\
$T_0$&       $2440949.89185 \pm 0.00088$ HJD&$T_0$&       $2440949.8967 \pm 0.0014$ HJD\\
$P_0$& $0.199684393 \pm (8.9 \times 10^{-8})$ d&$P_0$& $0.199683305 \pm (2.6 \times 10^{-7})$ d\\
$c_0$& $(6.06 \pm 0.17) \times 10^{-11}$ d&$c_0$& $(1.17 \pm 0.13) \times 10^{-10}$ d\\
					&&$d$ & $(-7.54 \pm 1.71) \times 10^{-16}$ d\\
$\chi^2$ (dof)&       168.42 (84)&$\chi^2$ (dof)&       136.37 (83)\\
Conf. level&$1.3 \times 10^{-7}$&Conf. level&$2.0 \times 10^{-4}$\\

\end{tabular}
\end{table}

\section*{Orbital Ephemeris}

The ASM arrival times were combined with those from earlier soft X-ray
observations and fit with both quadratic and cubic ephemeris models
(Figure \ref{res_plot}). The resulting best-fit parameters are shown in
Table \ref{ephem}. Neither quadratic nor cubic models provide a
statistically acceptable fit to all the data. Possibly the errors in
earlier data have been underestimated due to cycle-to-cycle variations
or other systematic effects \cite{kita95}. No single data set drives the bad
$\chi^2\/$.

The cubic term is still statistically significant, but the fitted
magnitude has decreased with time: from $\ddot{P} = 1.2 \times 10^{-10}$
\cite{vand89} to $4.1 \times 10^{-11}$ yr$^{-1}$ in this work. The wind
model predicts essentially zero $\ddot{P}$. Future ASM data should
provide a definitive test of the reality of the non-zero $\ddot{P}$.

\section*{OSSE Arrival Time Derivation}

Following the same procedure used for the ASM data, arrival times were
determined for the each of the 7 OSSE observations of Cyg X--3 from 1991
to 1994 using light curves of 2-min background-subtracted counting rates
$\sim 44$--130 \keV. The arrival times for the 7 observations were
compared with the values predicted using the soft X-ray ephemerides
(Table \ref{ephem}). The OSSE points fall systematically below the
predicted curves on the residual plot.

The OSSE phase minimum ($\sim 44$--130 \keV) is significantly
(6--7$\sigma$) earlier than the predicted soft X-ray (1--10 \keV)
minimum for both quadratic and cubic ephemerides ($\Delta t\sim -20 \pm
3$ min, $\Delta\phi \sim -0.065 \pm 0.010$). The data are consistent
with a constant offset over more than 3 years of observations.

\section*{Orbital light curve analysis}

\begin{figure}[tbh]
\centerline{\epsfig{file=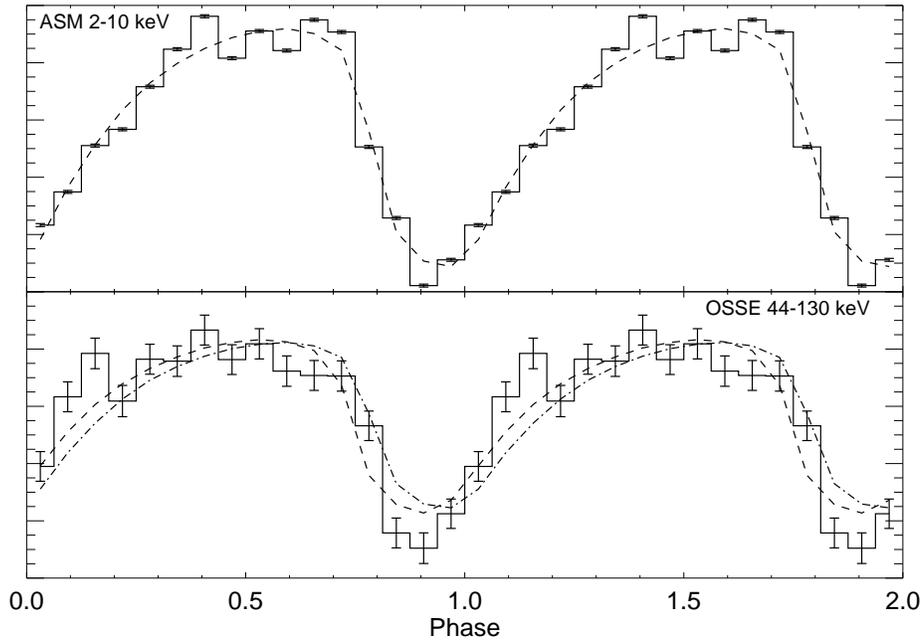,width=5in,angle=90}}
\caption{ASM (top) and OSSE (bottom) light curves with best-fit X-ray
templates overplotted (dashed line). The dashed-dotted line in the OSSE
plot shows the best fit to the OSSE data with the ASM phase. The shift
between the two energies is apparent.}  \label{lc}  \end{figure}

Light curves (Figure \ref{lc}) were produced from the ASM and OSSE data
using the quadratic ephemeris in Table \ref{ephem}. The plotted errors
in the ASM data reflect only the reported statistical errors. The light
curves were fit with the EXOSAT soft X-ray template \cite{vand89},
varying phase, amplitude, and DC intensity to determine the phase of
minimum and the consistency of the overall shape with the template.

Both the OSSE and ASM light curves are statistically {\em inconsistent}
with the canonical X-ray template \cite{vand89}. Qualitatively, however,
the template reasonably describes the ASM data. The large $\chi^2$ is
due to the fact that the actual observed variations in each bin are
larger that the statistical errors due to underlying source fluctuations. 

Evolution of the light curve shape with time is expected if the apparent
orbital period change is due to apsidal motion \cite{ghos81,elsn80}.
There is no evidence for this in the ASM observations.

The OSSE LC appears somewhat more symmetric than the template, with a
faster rise, but this is difficult to constrain with current data. The
differences in LC shape may contribute to the apparent phase shift with
energy.

\end{document}